\documentclass[noshowpacs,amsmath,
twocolumn,
superscriptaddress,
8pt
]{revtex4-1}
\usepackage{setspace}
\usepackage{amsmath}
\usepackage{float}
\usepackage{bm}
\usepackage{ulem}
\usepackage{graphicx}
\usepackage[nearskip,margin = 0pt]{subfig}

\usepackage{verbatim}
\usepackage{amsfonts}
\usepackage{braket}
\usepackage{amssymb}
\usepackage{upgreek}
\usepackage[colorlinks,linkcolor=blue,anchorcolor=blue,citecolor=blue,urlcolor=black]{hyperref}
\usepackage{epstopdf}
\usepackage{xcolor}
\usepackage{booktabs}
\usepackage{tabularx}
\usepackage{xtab}
\usepackage{changepage}
\usepackage{ragged2e}

\DeclareGraphicsExtensions{.pdf,.eps,.png,.jpg,.mps} 

\begin{document}
\title{Controlling the 2D magnetism of CrBr$\bf{_3}$ by van der Waals stacking engineering}

\author{Shiqi Yang$^{1,\star}$, Xiaolong Xu$^{2,\star,\dagger}$, Bo Han$^{3,4}$, Pingfan Gu$^{1}$, Roger Guzman$^{5}$, Yiwen Song$^{1}$, Zhongchong Lin$^{1}$, Peng Gao$^{3,4,6,7}$, Wu Zhou$^{5}$, Jinbo Yang$^{1}$, Zuxin Chen$^{8,\dagger}$ and Yu Ye$^{1,7,9,10,\dagger}$\\
\vspace{6pt}
$^1$State Key Laboratory for Mesoscopic Physics and Frontiers Science Center for Nano-optoelectronics, School of Physics, Peking University, Beijing 100871, China\\
$^2$School of Integrated Circuits and Electronics, MIIT Key Laboratory for Low-Dimensional Quantum Structure and Devices, Beijing Institute of Technology, Beijing 100081, China\\
$^3$Electron Microscopy Laboratory, School of Physics, International Center for Quantum Materials, Peking University, Beijing 100871, China\\
$^4$International Center for Quantum Materials, School of Physics, Peking University, Beijing, 100871, China\\
$^{5}$School of Physical Sciences, University of Chinese Academy of Sciences, Beijing 100049, China\\
$^{6}$Interdisciplinary Institute of Light-Element Quantum Materials and Research Center for Light-Element Advanced Materials, Peking University, Beijing 100871, China\\
$^{7}$Collaborative Innovation Centre of Quantum Matter, Beijing 100871, China\\
$^8$School of Semiconductor Science and Technology, South China Normal University, Foshan 528225, China\\
$^9$Yangtze Delta Institute of Optoelectronics, Peking University, Nantong 226010 Jiangsu, China\\
$^{10}$Liaoning Academy of Materials, Shenyang 110167, China\\
\vspace{3pt}
$^{\star}$These authors contributed equally\\
$^{\dag}$Corresponding to: xuxiaolong@bit.edu.cn, chenzuxin@m.scnu.edu.cn, ye\_yu@pku.edu.cn
}

\begin{abstract}
\begin{adjustwidth}{-2cm}{0cm}
\textbf{The manipulation of two-dimensional (2D) magnetic order is of significant importance to facilitate future 2D magnets for low-power and high-speed spintronic devices. Van der Waals stacking engineering makes promises for controllable magnetism via interlayer magnetic coupling. However, directly examining the stacking order changes accompanying magnetic order transitions at the atomic scale and preparing device-ready 2D magnets with controllable magnetic orders remain elusive. Here, we demonstrate effective control of interlayer stacking in exfoliated CrBr$_3$ via thermally assisted strain engineering. The stable interlayer ferromagnetic (FM), antiferromagnetic (AFM), and FM-AFM coexistent ground states confirmed by the magnetic circular dichroism measurements are realized. Combined with the first-principles calculations, the atomically-resolved imaging technique reveals the correlation between magnetic order and interlay stacking order in the CrBr$_3$ flakes unambiguously. A tunable exchange bias effect is obtained in the mixed phase of FM and AFM states. This work will introduce new magnetic properties by controlling the stacking order, and sequence of 2D magnets, providing ample opportunities for their application in spintronic devices.}
\end{adjustwidth}
\end{abstract}
\date{\today}
\maketitle

\noindent
Van der Waals materials with intrinsic magnetism have attracted tremendous attention due to their atomic-scale thickness that can be used to realize integrable and flexible magnetic devices\cite{2Dmagnetism-5,2Dmagnetism-4,2Dmagnetism-1,2Dmagnetism-2,2Dmagnetism-3}. Effective control of their magnetic orders is the real power with which we can fabricate all kinds of spintronic devices. In particular, when two-dimensional (2D) magnetism was first demonstrated\cite{2Dmagnetism-5}, it was surprising to find that the exfoliated thin CrI$_3$ behaves as an A-type antiferromagnetic (AFM) semiconductor with intralayer ferromagnetic (FM) coupling and interlayer AFM coupling, while the bulk CrI$_3$ crystal behaves as a ferromagnet at low temperatures. A large number of subsequent experiments and theories confirmed that this intriguing difference comes from the change of the interlayer magnetic coupling caused by the different stacking order in CrI$_3$ and CrCl$_3$\cite{pressure-1,pressure-2,CrCl3-TMR-Raman,stacking-nv,stacking-shg,stacking-2,stacking-3,stacking-4}. Specifically, CrI$_3$ crystal undergoes a transition from the monoclinic (M) phase to rhombohedral (R) phase below room temperature\cite{CrI3-CM,CrX31964,CrX3,CrI3-CM,cRx31952} and exhibit interlayer FM coupling at low temperature, while exfoliated thin layers of CrI$_3$ will maintain the monoclinic phase at low temperatures, which manifests as interlayer AFM coupling. Based on these understandings, strategies such as hydrostatic pressure have been proposed to effectively control the interlayer stacking order and thus their magnetic properties\cite{electric-1,electrical-2,electrical-3,pressure-1,pressure-2,stacking-nv,CrCl3-TMR-Raman,strain-hole-CrI3}, and recently promote the extensive development of moir\'e magnetism\cite{moire-1,moire-2,moire-3,miore-4}. Nevertheless, the correlation between crystal structure and magnetic order is mainly characterized by optical means such as Raman spectroscopy and nonlinear optics\cite{pressure-1,pressure-2,CrCl3-TMR-Raman,stacking-shg}, and the verification of atomic resolution is still lacking. Surprisingly, as an isostructural material, both bulk and exfoliated CrBr$_3$ flakes have been demonstrated and widely used as perfect ferromagnetic semiconductors\cite{CrBr3doamin,CrBrmicromagnetometry,CrBrspinfluctuation,thinCrX3-TMR,CrBr-PL}. The remaining questions are (1) what is the essential difference of CrBr$_3$ compared with CrI$_3$ and CrCl$_3$, and (2) is there an effective method to control the magnetic order of CrBr$_3$?

\begin{figure*}[!tb]
\includegraphics[width=2\columnwidth]{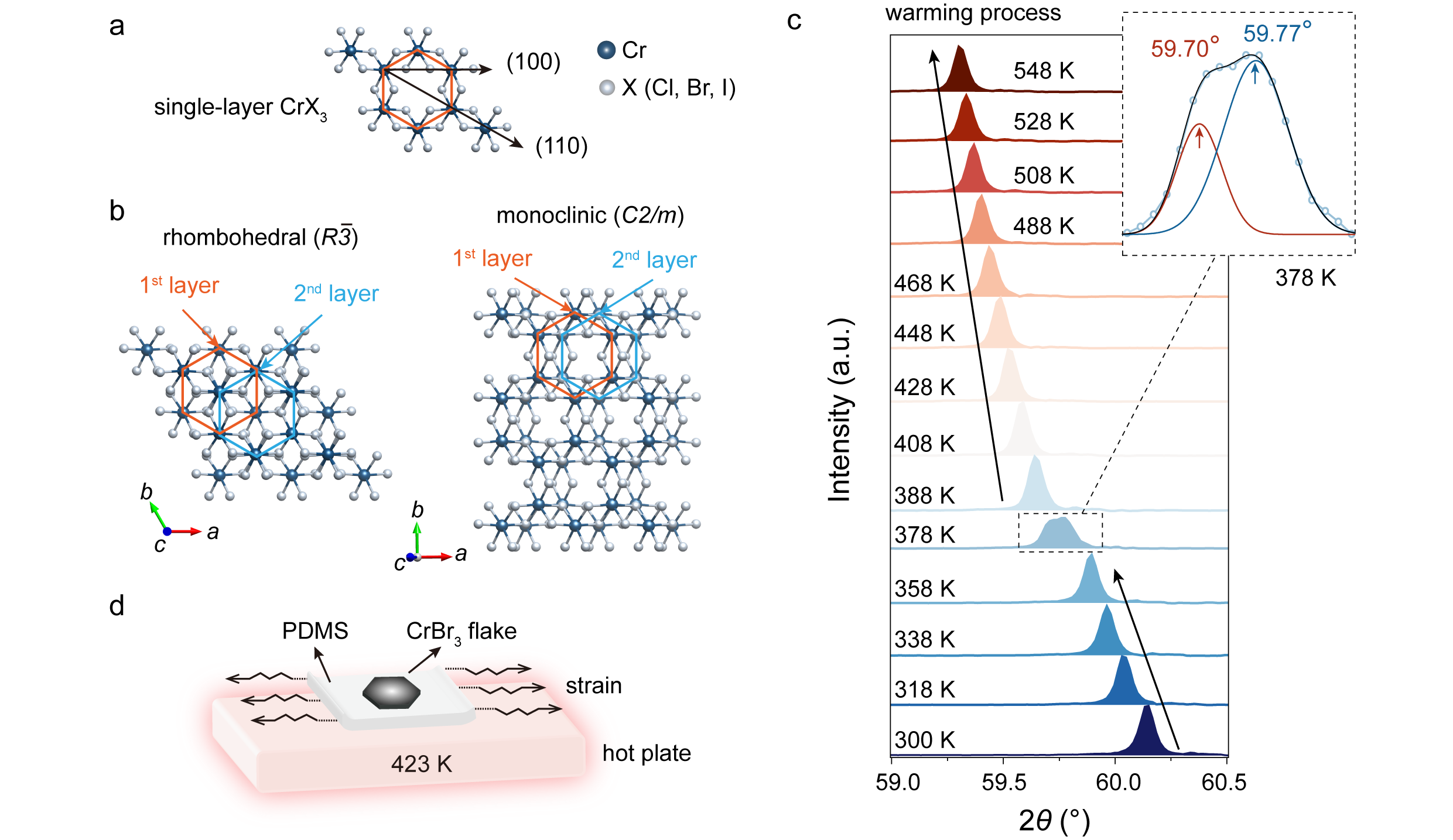}
    \captionsetup{singlelinecheck=off, justification = RaggedRight}
    \caption{\label{F1}\textbf{Structure phase transition of bulk $\bf{CrBr_3}$ crystals.}
     \textbf{a.} Atomic structure of single-layer CrX$_3$ (X=Cl, Br, I). Cr atoms (blue spheres) form a honeycomb lattice structure surrounded by six X atoms (gray spheres), which are octahedrally coordinated.
    \textbf{b.} Crystal structures of rhombohedral (R, left) and monoclinic (M, right) CrBr$_3$.
    \textbf{c.} Temperature-dependent (004) diffraction peak of CrBr$_3$ crystal measured by XRD during one arming process. The enlarged spectrum at 378 K shows a distinct two-peak behavior.
    \textbf{d.} Schematic illustration of thermal-assisted strain engineering exfoliated CrBr$_3$ flakes.
}
\end{figure*}

Unlike CrI$_3$ and CrCl$_3$ crystals that transition from M phase to R phase at low temperatures\cite{CrX31964,CrX3,CrI3-CM,cRx31952}, earlier work predicted that CrBr$_3$ would undergo this structural phase transition above room temperature\cite{CrBr1964}, but it has not been experimentally confirmed. Additionally, previous pioneering work verified the correlation between the interlayer magnetic coupling and the stacking order in molecular beam epitaxy-synthesized CrBr$_3$ bilayers by in situ spin-polarized scanning tunneling microscopy\cite{CrBrSTM}. Based on the better stability of CrBr$_3$ and the expected structural phase transition above room temperature\cite{CrBr-PL,CrBr1964,CrBrSTM,crIdegra}, it provides a unique platform for effectively controlling the magnetic order through van der Waals stacking engineering, which is crucial for the realization of related spintronic devices. Here, effective interlayer stacking control is achieved in exfoliated CrBr$_3$ flakes by thermally assisted strain engineering, and the correlation between van der Waals stacking order and magnetic order is verified at the atomic scale unambiguously. 

\begin{figure*}[tb]
   \includegraphics[width=2\columnwidth]{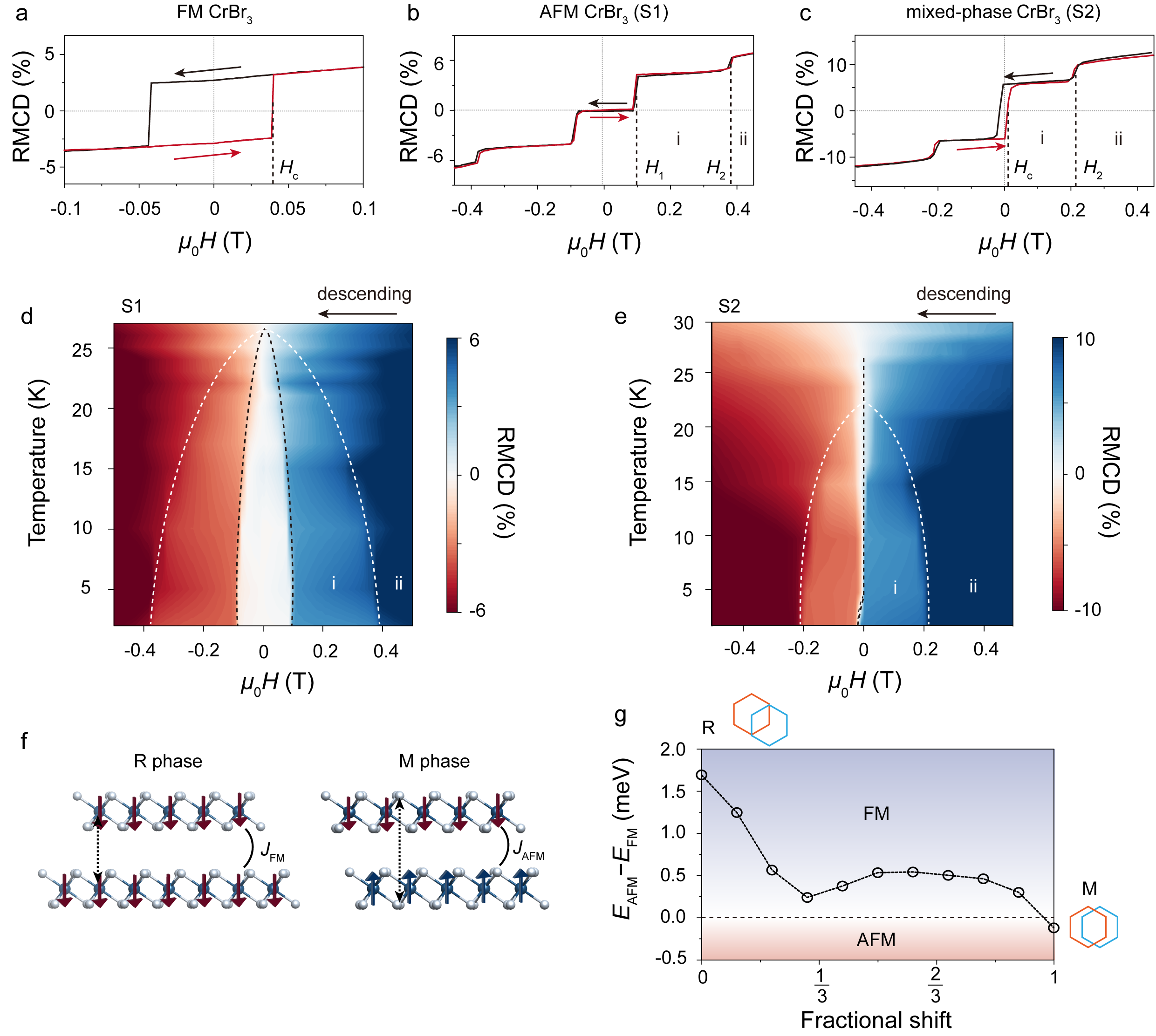}
   \captionsetup{singlelinecheck=off, justification = RaggedRight}
   \caption{\label{F2}\textbf{Stacking-related magnetism in exfoliated $\bf{CrBr_3}$ flakes.} 
    \textbf{a-c.} RMCD signal $versus$ magnetic field sweeping curves of the FM CrBr$_3$ sample (\textbf{a}) prepared at room temperature, and the AFM (\textbf{b}) and mixed-phase (\textbf{c}) CrBr$_3$ samples prepared by thermal-assisted strain engineering process. The spin-flip fields are expressed as $H_{\rm{c}}$ for FM CrBr$_3$, $H_{\rm{1}}$ and $H_{\rm{2}}$ for AFM CrBr$_3$.
    \textbf{d-e.} For S1 (\textbf{d}) and S2 (\textbf{e}) CrBr$_3$ samples, the RMCD signal plot in the parameter space of temperature and $\mu_0H$ under descending field sweep. 
    \textbf{f.} Schematic diagram of the magnetic ground state of 2L R and M phase CrBr$_3$ obtained by calculation.
    \textbf{g.} Calculated energy difference between AFM and FM states during 2L CrBr$_3$ shift from R phase to M phase.
  }
\end{figure*}

We start with the crystallographic structure of bulk CrBr$_3$ crystals. The single-crystal X-ray diffraction (SC-XRD) spectrum of the as-grown CrBr$_3$ crystal collected at 273 K (Supplementary Table 1) gives an R structure ($R\bar{3}$, $a = 6.306$ \AA, $c = 18.372$ \AA), different from the M phase of CrI$_3$ and CrCl$_3$ crystals at room temperature\cite{CrX31964,CrX3,CrI3-CM}. In a single-layer CrBr$_3$, Cr atoms form a honeycomb structure surrounded by six octahedrally coordinated Br atoms (Fig. \ref{F1}a). Sliding along the high-symmetry [$1\bar{1}0$] direction and stacking these monolayers will yield R phase of CrBr$_3$ (left panel of Fig. \ref{F1}b). Magnetic measurements of CrBr$_3$ crystal reveals long-range ferromagnetism with a Curie temperature ($T_{\rm{c}}$) of about 33 K and exhibit an out-of-plane magnetic anisotropy with a relatively low spin-flip field ($\sim$ 0.25 T) (Supplementary Fig. S1), indicating that the room-temperature (RT) CrBr$_3$ R phase corresponds to interlayer FM coupling at temperatures below its $T_{\rm{c}}$.

\begin{figure*}[tb]
	\includegraphics[width=2\columnwidth]{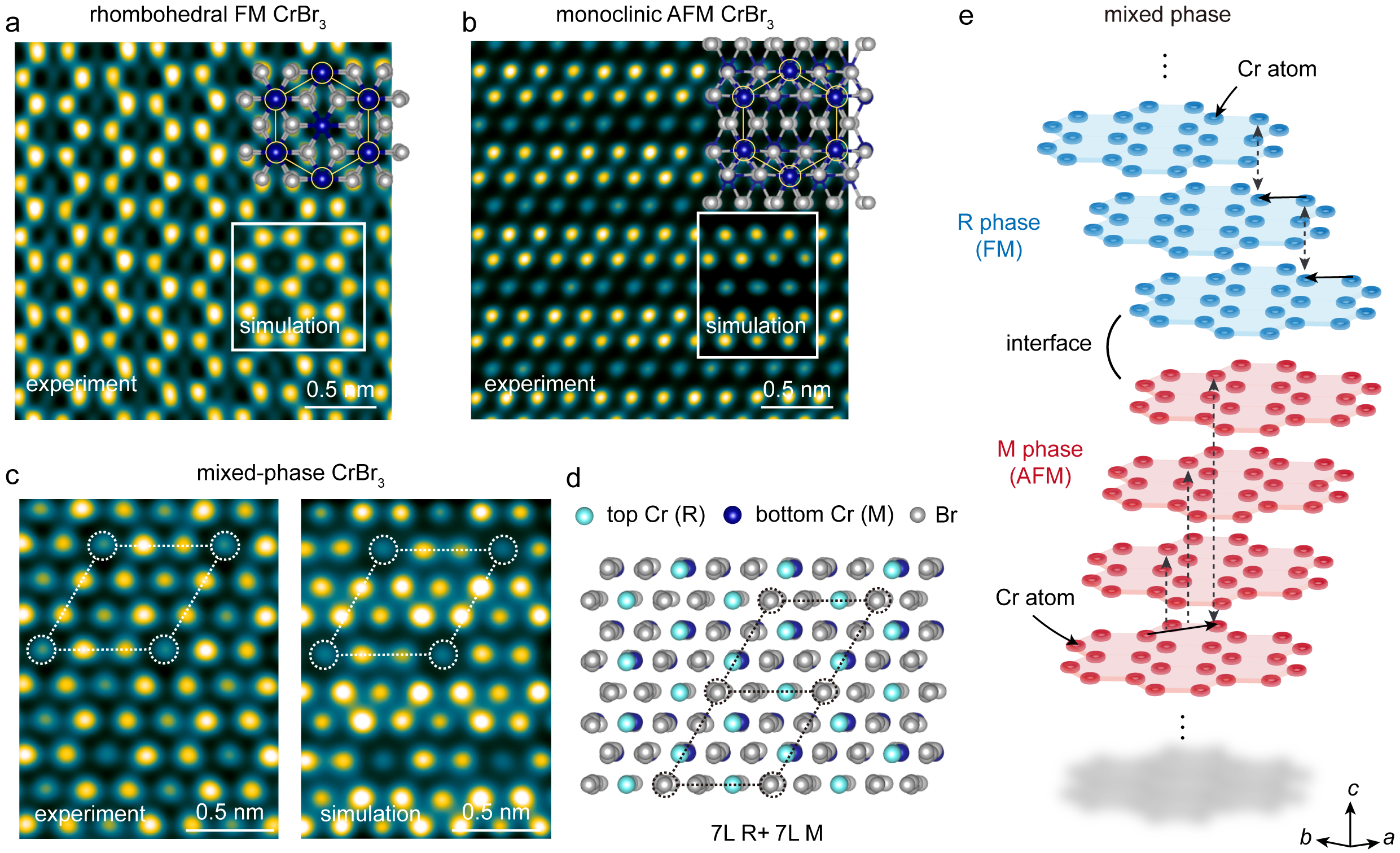}
	\captionsetup{singlelinecheck=off, justification = RaggedRight}
    \caption{\label{F3}\textbf{STEM characterizations and corresponding simulations of FM, AFM, and mixed-phase $\bf{CrBr_3}$}. 
    \textbf{a-b.} HAADF-STEM images of FM (\textbf{a}) and AFM (\textbf{b}) CrBr$_3$ reveal the rhombohedral (\textbf{a}) and monoclinic (\textbf{b}) structures, in good agreement with the corresponding simulated images inset. The stripe contrast in the monoclinic structure is due to the arrangement of Cr-Br and Br hollow columns.
    \textbf{c.} HAAD-STEM image of a mixed-phase CrBr$_3$, revealing a new periodic structure whose unit cell is marked by the white dashed lines. The right panel shows the simulation results of the vertically stacked 7L R phase and 7L M phase. 
    \textbf{d.} The atomic arrangement of Cr and Br atoms in mixed-phase CrBr$_3$ utilized for HAADF-STEM simulations. 
    \textbf{e.} Schematic of the atomic model illustrating the vertical stacking of R and M phases. 
    }
\end{figure*}

To fully understand the evolution of the crystal structure with temperature, especially to identify possible structural phase transition, powder XRD measurements were performed on CrBr$_3$ crystals between 300 K and 548 K. Figure \ref{F1}c shows the temperature-dependent XRD measurements during one warming process. Near 378 K, the diffraction angle is clearly split into two peaks at 59.70° and 59.77°, which are almost linearly red-shifted with increasing temperature on each side of 378 K, respectively, indicative of a temperature-induced structure phase transition. The significant thermal hysteresis during cooling and warming processes demonstrates the first-order nature of the transition (Supplementary Fig. 2). Differential scanning calorimetry measurement further confirmed the phase transition with an endothermic peak at an exact temperature of 373.7 K (Supplementary Fig. 2). Is the HT phase the M phase (right panel of Fig. \ref{F1}b, CrBr$_3$ monolayers slide along the [100] direction and stack, $C2/m$) as we expected, and what is its magnetic order? However, once cooling down, the HT phase will transform back to the RT R phase, which hinders us from revealing the crystal structure and magnetic properties of the HT phase. 

Given the knowledge from extensive research on chromium trihalides\cite{pressure-1,pressure-2,CrCl3-TMR-Raman,stacking-nv,stacking-shg,stacking-2,stacking-3,stacking-4,CrI3-CM,CrX3,CrX31964,cRx31952}, we propose a thermally assisted strain engineering approach to fix the CrBr$_3$ HT phase (Fig. \ref{F1}d). In an argon-filled glove box, the CrBr$_3$ flakes were exfoliated onto the polydimethylsiloxane (PDMS) substrate, and then the PDMS was carefully stretched several times on a hot plate set at 423 K. The stretched flakes subsequently transferred onto Si/SiO$_2$ substrates also display the same magnetic properties as those on PDMS substrates (Supplementary Fig. 3). The magnetic properties of few-layer CrBr$_3$ flakes were probed using reflective magnetic circular dichroism (RMCD) microscopy (see Methods for details). CrBr$_3$ flakes exfoliated and stretched at room temperature were first fabricated for comparison (Supplementary Fig. 4). The RMCD signal \textit{versus} magnetic field of a $\sim$10 nm flake (Fig. \ref{F2}a) shows a distinct hysteresis loop with a spin-flip field of 0.04 T at 2 K, confirming its FM nature with $T_{\rm{c}}$ of about 33 K (Supplementary Fig. 5). When the thickness increases, the rectangular hysteresis vanishes and multiple magnetic transitions occur, which is in accordance with previous reports\cite{CrBr-PL,CrBr3doamin}.

\begin{figure*}[tb]
	\includegraphics[width=2\columnwidth]{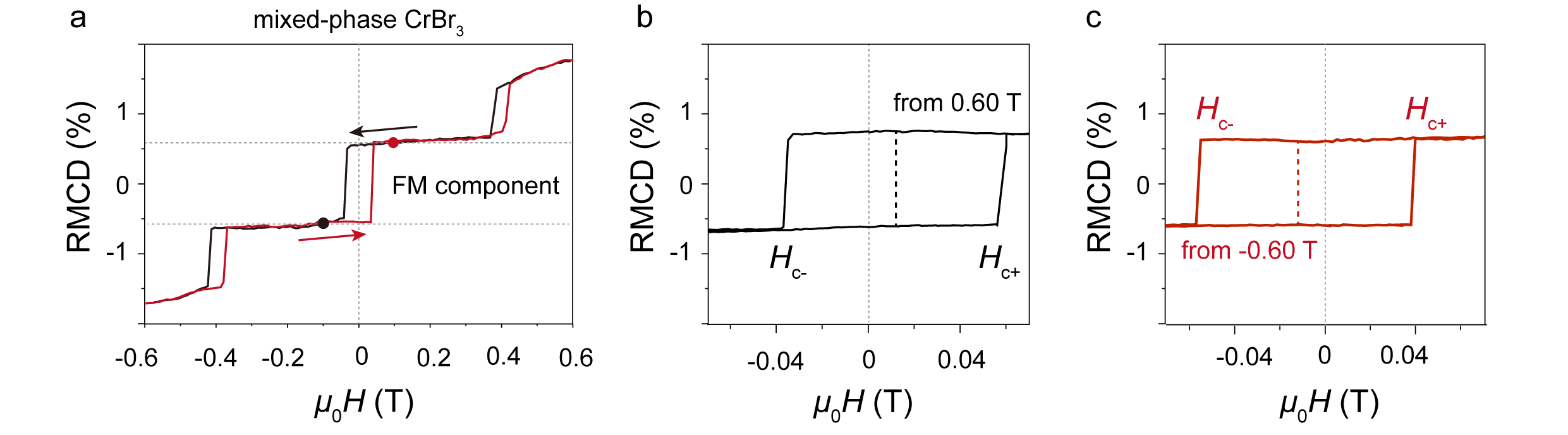}
    \captionsetup{singlelinecheck=off, justification = RaggedRight}
	\caption{\label{F4}\textbf{Exchange bias effects in the mixed-phase $\bf{CrBr_3}$ at 2 K}. \textbf{a.} RMCD signal \textit{versus} $\mu_0H$ for a mixed-phase CrBr$_3$ sample (S3) at 2 K. Solid circles define the magnetic field sweep range in the exchange bias study. 
    \textbf{b-c.} The RMCD signals \textit{versus} $\mu_0H$ swept within $\pm$0.10 T after applying a saturation field of 0.60 T (\textbf{b}) or $-$0.60 T (\textbf{c}), yielding opposite exchange bias fields of +12 mT (\textbf{b}) and $-$12 mT (\textbf{c}).
    }
\end{figure*}

In the thermally assisted stretched CrBr$_3$ flakes, we observe drastically different behaviors from the FM flakes, and we can classify them into two categories. First, in the S1-class samples (Fig. \ref{F2}b), a plateau with zero RMCD signal is shown within $\pm$0.10 T, indicating typical AFM behavior. Then, the forced FM state is reached through a two-step spin-flip transition around 0.10 T and 0.40 T ($H_{\rm{1}}$ and $H_{\rm{2}}$, marked by the dashed lines in Fig. \ref{F2}b), which resembles the even-layer A-type AFM CrI$_3$ but with weaker interlayer exchange coupling\cite{moire-1}. We speculate that the pure AFM state of CrBr$_3$ is also from the M phase, which will be verified later by scanning transmission electron microscopy (STEM) characterizations. As the temperature increase, the spin-flip fields decrease continuously and vanish around 27 K when the flake becomes paramagnetic (Fig. \ref{F2}d and Supplementary Fig. 6). The critical temperature of the AFM CrBr$_3$ flake (N\'eel temperature, $T_{\rm{N}}$) is slightly lower than that of the FM flake ($T_{\rm{c}}$ of 33 K). Second, apart from the pure AFM samples, the S2-class samples show obvious residual magnetic moment near zero magnetic field (indicating the presence of an FM component), and then undergo another spin-flip ($H_{\rm{2}}$) to a forced FM state under high fields (Fig. \ref{F2}c). The magnetic configuration of the S2 sample is different from that of the uncompensated odd-layer A-type AFM sample but exhibits the coexistence of FM and AFM components because (1) All S2-class samples with different thicknesses exhibit distinct FM signals around 0 T; (2) The spin-flip fields of $H_{\rm{c}}$ and $H_{\rm{2}}$, and the resulting changes in the RMCD signals, do not show obvious thickness dependence (Supplementary Fig. 7); (3) $H_{\rm{c}}$ and $H_{\rm{2}}$ are distributed in wide ranges of 0--0.10 T and 0.27--0.42 T, respectively (Supplementary Fig. 7), indicating that the FM and AFM components are coupled to each other; (4) The AFM features seem to disappear at a lower temperature than the FM features (Fig. \ref{F2}e), consistent with the fact that $T_{\rm{N}}$ $<$ $ T_{\rm{c}}$ for pure AFM and FM samples. Based on the above experimental observations, the coexistence of FM and AFM components in the S2 sample might come from an incomplete phase transition, resulting in the coexistence of R and M phases (mixed phase).

First-principle calculations in a 2L CrBr$_3$ link crystal stacking and magnetic order. The transition from R to M phase can be regarded as the sliding of the $2^{nd}$ layer of CrBr$_3$ in the R phase. Density functional theory (DFT) calculations reveal the energy difference between the AFM state and the FM state during shifting from the R phase to the M phase (Fig. \ref{F2}f). In the R phase, the FM state is more stable with an energy of 1.7 meV/Cr lower than the AFM state. As it approaches the M phase, the energy difference decreases and eventually crosses zero and becomes negative, indicating that the M phase is in the AFM ground state (Fig. \ref{F2}g). This is consistent with our RMCD measurements of CrBr$_3$ RT FM phase and HT AFM phase. 

To experimentally confirm the atomic structures of AFM and mixed-phase CrBr$_3$ samples and verify the correlation between magnetism and stacking order, we transferred representative FM, AFM, and mixed-phase samples after RMCD measurements onto copper microgrids for STEM characterizations. The high-angle annular dark-field (HAADF)-STEM image of the FM sample is highly consistent with the corresponding simulation results (Fig. \ref{F3}a), and shows a uniform rhombohedral structure on a large scale (Supplementary Fig. 8), which agrees with the SC-XRD measurement results. For the AFM sample, the HAADF-STEM image shows an obvious anisotropic crystal structure (Fig. \ref{F3}b). Comparison with the simulated monoclinic structure confirms that the stripe contrast is caused by the arrangement of Cr-Br (bright stripes) and Br hollow (dark stripes) atomic columns in the $ab$ plane. RMCD measurements, and DFT calculations combined with STEM characterizations collectively confirmed that our thermally assisted strain process can engineer the van der Waals stacking in CrBr$_3$, and thus effectively control its 2D magnetic properties.

Furthermore, HAADF-STEM of mixed-phase CrBr$_3$ samples exhibit a series of complex new periodic structures (Supplementary Fig. 8). Take the zoomed-in image of Fig. \ref{F3}c as an example, a unit cell (marked by the white dashed lines) contains dark spots on the four corners and irregular distributed bright and less bright spots inside. The simulation result of the 7L R phase stacked vertically on the 7L M phase (right panel in Fig. \ref{F3}c) agrees well with the experimental result. Fig. \ref{F3}d-e show schematic diagrams of the atomic model in top and side views, illustrating the vertical stacking of coexisted R (FM state) and M (AFM state) phases in mixed-phase CrBr$_3$. Small discrepancies between the simulation and experiment suggest that the exact vertical stacking in this structure may be more than a simple superposition of R and M phases, but complex combinations of the two components.

The coexistence of FM and AFM orders in the mixed-phase $\rm{CrBr_3}$ sample produces abundant AFM/FM interfaces (vertical and/or in-plane), which provide a platform for studying the exchange interactions established at the interfaces. Sweeping in a large magnetic field range of $\pm$0.60 T, the FM component encounters a time-reversal symmetric AFM environment when flipping from up to down and down to up (Fig. \ref{F4}a). The FM component within the minor hysteresis loop of $\pm$0.10 T experiences a constant AFM environment, which would generate an unprecedented exchange bias (EB) effect due to the coupling between FM and AFM components. After being historically polarized by a positive saturation magnetic field of 0.60 T, the $H_{\rm{c+}}$ of the minor FM hysteresis loop shifts to the right, showing a positive EB with an exchange field of $+$12 mT (Fig. \ref{F4}b). In contrast, after being polarized by a negative saturation magnetic field of $-$0.60 T, the $H_{\rm{c-}}$ of the minor FM hysteresis loop shifts to the left, exhibiting a negative EB with an exchange field of $-$12 mT. The direction of this EB can be tuned by the historical polarization field, and the EB is rather stable under multiple back-and-forth magnetic field sweeps (Supplementary Fig. 9), suggesting that the mixed-phase CrBr$_3$ is an ideal platform for exploring interface physics and developing novel van der Waals magnet-based spintronic devices.

In summary, we experimentally realized effective control of interlayer magnetic coupling in exfoliated CrBr$_3$ flakes by a thermally assisted strain engineering approach and comprehensively demonstrated the correlation between atomically resolved stacking order and magnetism. In addition to the induction of pure FM and AFM magnetic ground states CrBr$_3$ flakes, we also reported mixed-phase CrBr$_3$ composed of FM and AFM components, leading to a tunable exchange bias effect. The precise interfaces in the vertical stacking and in-plane connection should be further examined to better understand the mechanism and application of the exchange bias effect, which may require cryo-focused ion beam (cryo-FIB) milling and in-situ STEM characterization. Our work broadens 2D magnetic material systems for studying and manipulating magnetic couplings and related physical properties\cite{CrBrspinfluctuation,CrBrmicromagnetometry,thinCrX3-TMR}, making them promising candidates for next-generation spintronic devices.

\bigskip
\noindent \textbf{\large Methods}\\
\noindent \textbf{Crystal synthesis and characterization.}
CrBr$_3$ single crystals were prepared by the chemical vapor transport (CVT) method\cite{CrBrgrowth}. High-purity Cr (28.8 mg, Alfa, 99.996 $\%$) and TeBr4 (371.2 mg, Alfa, 99.999 $\%$) were mixed (molar ratio of 1:1.5) and then sealed in a silica tube under vacuum. Thereafter, the evacuated silica tube was placed in a two-zone tubular furnace. Crystal growth was carried out for 5 days under a temperature gradient from 750 °C to 450 °C, using a heating/cooling rate of 1 °C/min. Temperature-dependent power XRD measurements were performed using Cu-K radiation ($\lambda$ = 1.5418 \AA). The magnetic properties of CrBr$_3$ crystals were measured using the physical properties measurement system (PPMS) produced by Quantum Design.

\noindent \textbf{Magneto-optical measurements.}
RMCD measurements were performed based on the Attocube closed-cycle cryostat (attoDRY2100) with a temperature down to 1.6 K and an out-of-plane magnetic field up to 9 T. The detailed setup and measurement have been described in our previous work. 

\noindent \textbf{STEM measurements.}
Atomic-resolution HAADF-STEM images were recorded using an aberration-corrected Titan Themis G2 microscope operating at 300 kV. The convergence semi-angle is 30 mrad, and the collection angle is 39-200 mrad.

\noindent \textbf{Density functional theory calculations.}
Our density functional theory (DFT) calculations were carried out using generalized gradient approximation and projector augmented wave methods, as implemented in the Vienna ab initio simulation package (VASP). The uniform k mesh of 13$×$13$×$1 was adopted for integration over the Brillouin zone (BZ). A plane-wave cutoff energy of 450 eV and a vacuum region larger than 15 \AA was used during the structural relaxations, and the residual force per atom in the optimized structures was less than 1 meV/\AA. We used the optB86b functional for structural-related calculations and the PBEsol function for energy comparisons among different magnetic configurations. The on-site Coulomb interaction to the Cr d orbitals had $U$ and $J$ values of 3.9 eV and 1.1 eV, respectively, as revealed by a linear response method and comparison with the experimental results.

\normalem
\bibliographystyle{naturemag}
\bibliography{ref}

\begin{thebibliography}{10}
\expandafter\ifx\csname url\endcsname\relax
  \def\url#1{\texttt{#1}}\fi
\expandafter\ifx\csname urlprefix\endcsname\relax\def\urlprefix{URL }\fi
\providecommand{\bibinfo}[2]{#2}
\providecommand{\eprint}[2][]{\url{#2}}

\bibitem{2Dmagnetism-5}
\bibinfo{author}{Huang, B.} \emph{et~al.}
\newblock \bibinfo{title}{Layer-dependent ferromagnetism in a van der waals
  crystal down to the monolayer limit}.
\newblock \emph{\bibinfo{journal}{Nature}} \textbf{\bibinfo{volume}{546}},
  \bibinfo{pages}{270--273} (\bibinfo{year}{2017}).

\bibitem{2Dmagnetism-4}
\bibinfo{author}{Gong, C.} \emph{et~al.}
\newblock \bibinfo{title}{Discovery of intrinsic ferromagnetism in
  two-dimensional van der waals crystals}.
\newblock \emph{\bibinfo{journal}{Nature}} \textbf{\bibinfo{volume}{546}},
  \bibinfo{pages}{265--269} (\bibinfo{year}{2017}).

\bibitem{2Dmagnetism-1}
\bibinfo{author}{Burch, K.~S.}, \bibinfo{author}{Mandrus, D.} \&
  \bibinfo{author}{Park, J.-G.}
\newblock \bibinfo{title}{Magnetism in two-dimensional van der waals
  materials}.
\newblock \emph{\bibinfo{journal}{Nature}} \textbf{\bibinfo{volume}{563}},
  \bibinfo{pages}{47--52} (\bibinfo{year}{2018}).

\bibitem{2Dmagnetism-2}
\bibinfo{author}{Gong, C.} \& \bibinfo{author}{Zhang, X.}
\newblock \bibinfo{title}{Two-dimensional magnetic crystals and emergent
  heterostructure devices}.
\newblock \emph{\bibinfo{journal}{Science}} \textbf{\bibinfo{volume}{363}},
  \bibinfo{pages}{eaav4450} (\bibinfo{year}{2019}).

\bibitem{2Dmagnetism-3}
\bibinfo{author}{Gibertini, M.}, \bibinfo{author}{Koperski, M.},
  \bibinfo{author}{Morpurgo, A.~F.} \& \bibinfo{author}{Novoselov, K.~S.}
\newblock \bibinfo{title}{Magnetic 2{D} materials and heterostructures}.
\newblock \emph{\bibinfo{journal}{Nature Nanotechnology}}
  \textbf{\bibinfo{volume}{14}}, \bibinfo{pages}{408--419}
  (\bibinfo{year}{2019}).

\bibitem{pressure-1}
\bibinfo{author}{Song, T.} \emph{et~al.}
\newblock \bibinfo{title}{Switching 2{D} magnetic states via pressure tuning of
  layer stacking}.
\newblock \emph{\bibinfo{journal}{Nature Materials}}
  \textbf{\bibinfo{volume}{18}}, \bibinfo{pages}{1298--1302}
  (\bibinfo{year}{2019}).

\bibitem{pressure-2}
\bibinfo{author}{Li, T.} \emph{et~al.}
\newblock \bibinfo{title}{Pressure-controlled interlayer magnetism in
  atomically thin {CrI$_3$}}.
\newblock \emph{\bibinfo{journal}{Nature Materials}}
  \textbf{\bibinfo{volume}{18}}, \bibinfo{pages}{1303--1308}
  (\bibinfo{year}{2019}).

\bibitem{CrCl3-TMR-Raman}
\bibinfo{author}{Klein, D.~R.} \emph{et~al.}
\newblock \bibinfo{title}{Enhancement of interlayer exchange in an ultrathin
  two-dimensional magnet}.
\newblock \emph{\bibinfo{journal}{Nature Physics}}
  \textbf{\bibinfo{volume}{15}}, \bibinfo{pages}{1255--1260}
  (\bibinfo{year}{2019}).

\bibitem{stacking-nv}
\bibinfo{author}{Thiel, L.} \emph{et~al.}
\newblock \bibinfo{title}{Probing magnetism in 2{D} materials at the nanoscale
  with single-spin microscopy}.
\newblock \emph{\bibinfo{journal}{Science}} \textbf{\bibinfo{volume}{364}},
  \bibinfo{pages}{973--976} (\bibinfo{year}{2019}).

\bibitem{stacking-shg}
\bibinfo{author}{Sun, Z.} \emph{et~al.}
\newblock \bibinfo{title}{Giant nonreciprocal second-harmonic generation from
  antiferromagnetic bilayer {CrI$_3$}}.
\newblock \emph{\bibinfo{journal}{Nature}} \textbf{\bibinfo{volume}{572}},
  \bibinfo{pages}{497--501} (\bibinfo{year}{2019}).

\bibitem{stacking-2}
\bibinfo{author}{Jiang, P.} \emph{et~al.}
\newblock \bibinfo{title}{Stacking tunable interlayer magnetism in bilayer
  {CrI$_3$}}.
\newblock \emph{\bibinfo{journal}{Physical Review B}}
  \textbf{\bibinfo{volume}{99}}, \bibinfo{pages}{144401}
  (\bibinfo{year}{2019}).

\bibitem{stacking-3}
\bibinfo{author}{Soriano, D.}, \bibinfo{author}{Cardoso, C.} \&
  \bibinfo{author}{Fern{\'a}ndez-Rossier, J.}
\newblock \bibinfo{title}{Interplay between interlayer exchange and stacking in
  {CrI$_3$} bilayers}.
\newblock \emph{\bibinfo{journal}{Solid State Communications}}
  \textbf{\bibinfo{volume}{299}}, \bibinfo{pages}{113662}
  (\bibinfo{year}{2019}).

\bibitem{stacking-4}
\bibinfo{author}{Jang, S.~W.}, \bibinfo{author}{Jeong, M.~Y.},
  \bibinfo{author}{Yoon, H.}, \bibinfo{author}{Ryee, S.} \&
  \bibinfo{author}{Han, M.~J.}
\newblock \bibinfo{title}{Microscopic understanding of magnetic interactions in
  bilayer {CrI$_3$}}.
\newblock \emph{\bibinfo{journal}{Physical Review Materials}}
  \textbf{\bibinfo{volume}{3}}, \bibinfo{pages}{031001} (\bibinfo{year}{2019}).

\bibitem{CrI3-CM}
\bibinfo{author}{McGuire, M.~A.}, \bibinfo{author}{Dixit, H.},
  \bibinfo{author}{Cooper, V.~R.} \& \bibinfo{author}{Sales, B.~C.}
\newblock \bibinfo{title}{Coupling of crystal structure and magnetism in the
  layered, ferromagnetic insulator {CrI$_3$}}.
\newblock \emph{\bibinfo{journal}{Chemistry of Materials}}
  \textbf{\bibinfo{volume}{27}}, \bibinfo{pages}{612--620}
  (\bibinfo{year}{2015}).

\bibitem{CrX31964}
\bibinfo{author}{Morosin, B.} \& \bibinfo{author}{Narath, A.}
\newblock \bibinfo{title}{X-ray diffraction and nuclear quadrupole resonance
  studies of chromium trichloride}.
\newblock \emph{\bibinfo{journal}{The Journal of Chemical Physics}}
  \textbf{\bibinfo{volume}{40}}, \bibinfo{pages}{1958--1967}
  (\bibinfo{year}{1964}).

\bibitem{CrX3}
\bibinfo{author}{McGuire, M.~A.}
\newblock \bibinfo{title}{Crystal and magnetic structures in layered,
  transition metal dihalides and trihalides}.
\newblock \emph{\bibinfo{journal}{Crystals}} \textbf{\bibinfo{volume}{7}},
  \bibinfo{pages}{121} (\bibinfo{year}{2017}).

\bibitem{cRx31952}
\bibinfo{author}{Handy, L.} \& \bibinfo{author}{Gregory, N.}
\newblock \bibinfo{title}{Structural properties of chromium (iii) iodide and
  some chromium (iii) mixed halides}.
\newblock \emph{\bibinfo{journal}{Journal of the American Chemical Society}}
  \textbf{\bibinfo{volume}{74}}, \bibinfo{pages}{891--893}
  (\bibinfo{year}{1952}).

\bibitem{electric-1}
\bibinfo{author}{Jiang, S.}, \bibinfo{author}{Shan, J.} \&
  \bibinfo{author}{Mak, K.~F.}
\newblock \bibinfo{title}{Electric-field switching of two-dimensional van der
  waals magnets}.
\newblock \emph{\bibinfo{journal}{Nature Materials}}
  \textbf{\bibinfo{volume}{17}}, \bibinfo{pages}{406--410}
  (\bibinfo{year}{2018}).

\bibitem{electrical-2}
\bibinfo{author}{Huang, B.} \emph{et~al.}
\newblock \bibinfo{title}{Electrical control of 2{D} magnetism in bilayer
  {CrI$_3$}}.
\newblock \emph{\bibinfo{journal}{Nature Nanotechnology}}
  \textbf{\bibinfo{volume}{13}}, \bibinfo{pages}{544--548}
  (\bibinfo{year}{2018}).

\bibitem{electrical-3}
\bibinfo{author}{Jiang, S.}, \bibinfo{author}{Li, L.}, \bibinfo{author}{Wang,
  Z.}, \bibinfo{author}{Mak, K.~F.} \& \bibinfo{author}{Shan, J.}
\newblock \bibinfo{title}{Controlling magnetism in 2d cri3 by electrostatic
  doping}.
\newblock \emph{\bibinfo{journal}{Nature Nanotechnology}}
  \textbf{\bibinfo{volume}{13}}, \bibinfo{pages}{549--553}
  (\bibinfo{year}{2018}).

\bibitem{strain-hole-CrI3}
\bibinfo{author}{Jiang, S.}, \bibinfo{author}{Xie, H.}, \bibinfo{author}{Shan,
  J.} \& \bibinfo{author}{Mak, K.~F.}
\newblock \bibinfo{title}{Exchange magnetostriction in two-dimensional
  antiferromagnets}.
\newblock \emph{\bibinfo{journal}{Nature Materials}}
  \textbf{\bibinfo{volume}{19}}, \bibinfo{pages}{1295--1299}
  (\bibinfo{year}{2020}).

\bibitem{moire-1}
\bibinfo{author}{Song, T.} \emph{et~al.}
\newblock \bibinfo{title}{Direct visualization of magnetic domains and
  moir{\'e} magnetism in twisted 2{D} magnets}.
\newblock \emph{\bibinfo{journal}{Science}} \textbf{\bibinfo{volume}{374}},
  \bibinfo{pages}{1140--1144} (\bibinfo{year}{2021}).

\bibitem{moire-2}
\bibinfo{author}{Xu, Y.} \emph{et~al.}
\newblock \bibinfo{title}{Coexisting ferromagnetic--antiferromagnetic state in
  twisted bilayer cri$_3$}.
\newblock \emph{\bibinfo{journal}{Nature Nanotechnology}}
  \textbf{\bibinfo{volume}{17}}, \bibinfo{pages}{143--147}
  (\bibinfo{year}{2022}).

\bibitem{moire-3}
\bibinfo{author}{Xie, H.} \emph{et~al.}
\newblock \bibinfo{title}{Evidence of non-collinear spin texture in magnetic
  moir{\'e} superlattices}.
\newblock \emph{\bibinfo{journal}{Nature Physics}} \bibinfo{pages}{1--6}
  (\bibinfo{year}{2023}).

\bibitem{miore-4}
\bibinfo{author}{Cheng, G.} \emph{et~al.}
\newblock \bibinfo{title}{Electrically tunable moir{\'e} magnetism in twisted
  double bilayers of chromium triiodide}.
\newblock \emph{\bibinfo{journal}{Nature Electronics}} \bibinfo{pages}{1--9}
  (\bibinfo{year}{2023}).

\bibitem{CrBr3doamin}
\bibinfo{author}{Sun, Q.-C.} \emph{et~al.}
\newblock \bibinfo{title}{Magnetic domains and domain wall pinning in
  atomically thin crbr$_3$ revealed by nanoscale imaging}.
\newblock \emph{\bibinfo{journal}{Nature communications}}
  \textbf{\bibinfo{volume}{12}}, \bibinfo{pages}{1989} (\bibinfo{year}{2021}).

\bibitem{CrBrmicromagnetometry}
\bibinfo{author}{Kim, M.} \emph{et~al.}
\newblock \bibinfo{title}{Micromagnetometry of two-dimensional ferromagnets}.
\newblock \emph{\bibinfo{journal}{Nature Electronics}}
  \textbf{\bibinfo{volume}{2}}, \bibinfo{pages}{457--463}
  (\bibinfo{year}{2019}).

\bibitem{CrBrspinfluctuation}
\bibinfo{author}{Jin, C.} \emph{et~al.}
\newblock \bibinfo{title}{Imaging and control of critical fluctuations in
  two-dimensional magnets}.
\newblock \emph{\bibinfo{journal}{Nature Materials}}
  \textbf{\bibinfo{volume}{19}}, \bibinfo{pages}{1290--1294}
  (\bibinfo{year}{2020}).

\bibitem{thinCrX3-TMR}
\bibinfo{author}{Kim, H.~H.} \emph{et~al.}
\newblock \bibinfo{title}{Evolution of interlayer and intralayer magnetism in
  three atomically thin chromium trihalides}.
\newblock \emph{\bibinfo{journal}{Proceedings of the National Academy of
  Sciences}} \textbf{\bibinfo{volume}{116}}, \bibinfo{pages}{11131--11136}
  (\bibinfo{year}{2019}).

\bibitem{CrBr-PL}
\bibinfo{author}{Zhang, Z.} \emph{et~al.}
\newblock \bibinfo{title}{Direct photoluminescence probing of ferromagnetism in
  monolayer two-dimensional {CrBr$_3$}}.
\newblock \emph{\bibinfo{journal}{Nano Letters}} \textbf{\bibinfo{volume}{19}},
  \bibinfo{pages}{3138--3142} (\bibinfo{year}{2019}).

\bibitem{CrBr1964}
\bibinfo{author}{Morosin, B.} \& \bibinfo{author}{Narath, A.}
\newblock \bibinfo{title}{X-ray diffraction and nuclear quadrupole resonance
  studies of chromium trichloride}.
\newblock \emph{\bibinfo{journal}{The Journal of Chemical Physics}}
  \textbf{\bibinfo{volume}{40}}, \bibinfo{pages}{1958--1967}
  (\bibinfo{year}{1964}).

\bibitem{CrBrSTM}
\bibinfo{author}{Chen, W.} \emph{et~al.}
\newblock \bibinfo{title}{Direct observation of van der waals
  stacking-dependent interlayer magnetism}.
\newblock \emph{\bibinfo{journal}{Science}} \textbf{\bibinfo{volume}{366}},
  \bibinfo{pages}{983--987} (\bibinfo{year}{2019}).

\bibitem{crIdegra}
\bibinfo{author}{Shcherbakov, D.} \emph{et~al.}
\newblock \bibinfo{title}{Raman spectroscopy, photocatalytic degradation, and
  stabilization of atomically thin chromium tri-iodide}.
\newblock \emph{\bibinfo{journal}{Nano letters}} \textbf{\bibinfo{volume}{18}},
  \bibinfo{pages}{4214--4219} (\bibinfo{year}{2018}).

\bibitem{CrBrgrowth}
\bibinfo{author}{Tartaglia, T.~A.} \emph{et~al.}
\newblock \bibinfo{title}{Accessing new magnetic regimes by tuning the ligand
  spin-orbit coupling in van der waals magnets}.
\newblock \emph{\bibinfo{journal}{Science Advances}}
  \textbf{\bibinfo{volume}{6}}, \bibinfo{pages}{eabb9379}
  (\bibinfo{year}{2020}).

\end{thebibliography}

\bigskip
\noindent \textbf{\large Data availability}\\
The source data generated in this study have been deposited in the database under the accession code (URL will be inserted before publication).

\bigskip
\noindent \textbf{\large Acknowledgement}\\
This work was supported by the National Key R\&D Program of China (No. 2022YFA1203902 and No. 2018YFA0306900), the National Natural Science Foundation of China (No. 12241401, No. 12250007 and No. 62274010), and Beijing Natural Science Foundation (No. JQ21018). We acknowledge the Electron Microscopy Laboratory of Peking University.

\bigskip
\noindent\textbf{\large Author contributions}\\
X.X., Y.Y., and S.Y. conceived the project, designed the experiments, analyzed the results, and draft the manuscript. Z.C. grew the CrBr$_3$ bulk crystals. B.H., Y.S., and P. Gao performed the STEM characterizations. R.G. and W.Z. simulated the HAADF-STEM results. P. Gu performed the first-principle calculations. Z.L. and J. Y. performed the temperature-dependent power XRD and magnetic property measurements of bulk CrBr$_3$ crystals. All authors discussed the results and contributed to the manuscript.

\bigskip
\noindent\textbf{\large Competing interests}\\
The authors declare no competing financial interests.

\bigskip
\noindent\textbf{\large Additional information}\\
Supplementary information is available for this paper at (URL need to be inserted).

\end{document}